\documentclass[showpacs,aps,pre,twocolumn]{revtex4-1}
\usepackage[ocgcolorlinks,colorlinks=true,linkcolor=blue,citecolor=blue,linktocpage=true]{hyperref}
\usepackage{amsmath}
\usepackage{amssymb}
\usepackage{graphicx}
\usepackage{ulem}
\usepackage{color,xcolor}

\newcommand{\ket}[1]{|#1\rangle}
\newcommand{\bra}[1]{\langle #1|}

\newcommand{\tr}{{\rm Tr}}

\begin{document}

\title{Entropy production and thermodynamic power of the squeezed thermal reservoir}
\author{Gonzalo Manzano,$^{1,2}$ Fernando Galve,$^2$ Roberta Zambrini,$^2$ and Juan M. R. Parrondo$^1$}
\affiliation{$^1$Departamento de F\'isica At\'omica, Molecular y Nuclear and GISC, Universidad Complutense Madrid, 28040 Madrid, Spain}
\affiliation{$^2$Instituto de F\'isica Interdisciplinar y Sistemas Complejos IFISC (CSIC-UIB), Campus Universitat Illes Balears, E-07122 Palma de Mallorca, Spain}
\date{\today}

\begin{abstract}
We analyze the entropy production and the maximal extractable work from a squeezed thermal reservoir. The nonequilibrium quantum 
nature of the reservoir induces an entropy transfer with a coherent contribution while modifying its thermal part, allowing 
work extraction from a single reservoir, as well as great improvements in power and efficiency for quantum heat engines. 
Introducing a modified quantum Otto cycle, our approach fully characterizes operational regimes forbidden in the standard case, 
such as refrigeration and work extraction at the same time, accompanied by efficiencies equal to unity.
\end{abstract}

\maketitle

\section{Introduction}

From the inception of equilibrium thermodynamics in the 19th century to the present, a great 
multidisciplinary effort has been devoted to its extension to far-from-equilibrium situations,  
some of the most important cornerstones being the development of thermodynamics at the stochastic 
level \cite{Sekimoto, SeifertReview} and its extension to the quantum regime \cite{CampisiReview, EspositoReview, KosloffReview}.
Furthermore, motivated by the success of quantum information theory and the increasing control in preparation 
and manipulation of quantum states, the last decade has experienced a growing interest in understanding the 
thermodynamic implications of quantum features, such as quantum measurement \cite{Hormoz,Jacobs, Brandner, Anders}, 
coherence \cite{Photocarnot, Photocell, Popescu, Aberg}, or quantum correlations \cite{Oppenheim, Zurek, Rio, Park, Brunner, Perarnau}. 
In this context, inspired by the breakthrough work on the photo-Carnot engine driven by quantum fuel proposed by 
Scully {\it et al.} \cite{Photocarnot}, different theoretical studies recently focused on the implications for work 
extraction introduced by nonequilibrium quantum reservoirs. In particular it has been shown that using coherent 
\cite{QuanPCE, ChineseII, Superradiant}, correlated \cite{Dillenschneider}, or squeezed thermal reservoirs 
\cite{ChineseHE, LutzHE, AdessoHE, Long}, power and efficiency of heat engines can be improved, even surpassing the 
Carnot bound. However a general framework providing a deeper understanding of such quantum nonequilibrium phenomena is 
still an open challenge \cite{Abah, Niedenzu}. 

In this paper we clarify the role of nonequilibrium quantum reservoirs via the analysis of entropy production,
one of the most fundamental concepts in non-equilibrium thermodynamics, which quantifies
the degree of irreversibility of a dynamical evolution \cite{PrigogineBook}.
For a quantum system relaxing in a thermal reservoir in equilibrium at inverse temperature $\beta = 1/k_B T$, 
it simply reads \cite{Lebowitz, Alicki, Deffner}
\begin{equation}\label{intro1}
\Sigma = \Delta S - \beta Q \geq 0 
\end{equation}
where $S = - \tr[\hat{\rho} \ln \hat{\rho}]$ denotes the von Neumann entropy of the system and $Q$ is the heat 
released from the reservoir. The positivity of the entropy production \eqref{intro1} is a particular case of the second law. 
However, in more general situations,
different processes others than heat flows may produce an exchange of entropy between the system and its 
surroundings, modifying (\ref{intro1}). We explicitly address such modifications and some of its 
counter-intuitive consequences for the case of a bosonic mode interacting with a 
squeezed thermal reservoir, giving a microscopic picture of the dynamical entropy exchange processes.
The maximum irreversible work cyclically extractable from a single squeezed reservoir is obtained.
Further, we discuss an Otto cycle which can operate as a heat engine converting
the heat entering from both reservoirs into work at unit efficiency, or as a refrigerator pumping energy from the cold to the hot 
reservoir while producing a positive amount of output work at the same time. Our results do not contradict the second law of thermodynamics, 
which is modified by the inclusion of squeezing as an available resource in the reservoir.
Squeezing is intimately related with Heisenberg's uncertainty principle, being the reduction of the 
variance of an observable with respect to the conjugate one \cite{QS}. Nowadays it constitutes a central 
tool in quantum information with several applications in quantum metrology, computation, cryptography, and 
imaging \cite{polzik}. Most commonly considered squeezed states are coherent, but also thermal ones have been 
largely studied \cite{Fearn,Knight}. Experimental realizations of squeezed thermal states range from micro-waves 
\cite{Yurke} to present squeezing of motional degrees of freedom in optomechanical oscillators \cite{schwab, Sillanpaa}.
 
\section{Thermodynamics of the squeezed thermal reservoir} \label{sec:thermo}
 
Consider a quantum system consisting of a single bosonic mode with 
Hamiltonian $\hat{H}_S = \hbar \omega \hat{a}^\dagger \hat{a}$, weakly
dissipating into a bosonic reservoir $\hat{H}_R = \sum_k \hbar \Omega_k \hat{b}_k^\dagger \hat{b}_k^{~}$, prepared in a squeezed thermal 
state at inverse temperature $\beta$ \cite{Breuer}.
The interaction between mode and reservoir $\hat{H}_{\rm int} = \sum_k i g_k (\hat{a}\, \hat{b}_k^\dagger - \hat{a}^\dagger \hat{b}_k^{~})$
yields an open system dynamics well described by the following Lindblad Master
Equation (LME) in interaction picture \cite{Walls, Scully} (see also Appendix \ref{app:collisional}):
\begin{eqnarray}\label{master-eq-fieldmode}
 \dot{\hat{\rho}}_S(t) = \mathcal{L}(\hat{\rho}_S(t)) = \sum_{i= \pm} \hat{R}_i \hat{\rho}_S(t) \hat{R}_i^\dagger - \frac{1}{2} \{ \hat{R}_i^\dagger \hat{R}_i, \hat{\rho}_S(t) \}, ~~~
\end{eqnarray}
where Lamb-Stark shifts have been neglected. The two Lindblad operators in (\ref{master-eq-fieldmode}) read 
$\hat{R}_{-} = \sqrt{\gamma (n_{\rm th} + 1)}~\hat{R}$ and $\hat{R}_{+} = \sqrt{\gamma n_{\rm th}}~\hat{R}^\dagger$, with 
$\hat{R} = \hat{a} \cosh(r) + \hat{a}^\dagger \sinh(r) e^{i \theta} = \hat{\mathcal{S}} \hat{a} \hat{\mathcal{S}}^\dagger$ and 
$\hat{\mathcal{S}}\equiv \exp(\frac{r}{2} (\hat{a}^2 e^{- i \theta} - \hat{a}^{\dagger 2} e^{i \theta}))$ denotes the 
unitary squeezing operator on the system mode $(r \geq 0$ and $ \theta \in [0, 2 \pi] )$, $\gamma$ is the spontaneous emission decay rate and $n_{\rm th}= (e^{\beta \hbar \omega} - 1)^{-1}$ 
the mean number of photons of frequency $\omega$ in a thermal reservoir at inverse temperature $\beta$. The operators $\hat{R}_{\mp}$, promote jumps associated 
with the correlated emission and adsorption of photons: $\hat{R}_{\mp} \hat{\mathcal{S}} \ket{n} \rightarrow  \hat{\mathcal{S}} \ket{n \mp 1}$, 
leading to a steady state solution, $ \mathcal{L}(\hat{\pi}_S)=0$, no longer diagonal in the $\hat{H}_S$ basis:
\begin{equation}\label{steady-state}
 \hat{\pi}_S = \hat{\mathcal{S}} \frac{e^{- \beta \hat{H}_S}}{Z} \hat{\mathcal{S}}^\dagger
\end{equation}
with $Z = \tr[e^{-\beta \hat{H}_S}]$. The squeezed thermal state $\hat{\pi}_S$ has the same entropy as the Gibbs state, but increased mean energy. 
A crucial property is that its variance in the quadrature $\hat{x}_{\theta/2} \equiv (\hat{a}^\dagger e^{i \theta/2} + \hat{a} e^{-i \theta/2})/\sqrt{2}$ 
has been squeezed by a factor $e^{-r}$, while the variance of the conjugate quadrature $\hat{p}_{\theta/2}$ ($[\hat{x}_{\theta/2}, \hat{p}_{\theta/2}]= i$)
is multiplied by  $e^{r}$. When turning to the Schr\"odinger picture, 
the steady state (\ref{steady-state}) acquires a time-dependent phase which has to be accounted for in applications.

The LME (\ref{master-eq-fieldmode}) describes relaxation of the mode to $\hat{\pi}_{S}$, 
the irreversibility of which is well captured by the so-called excess (or non-adiabatic) entropy production 
rate \cite{Spohn, EspositoTF, JordanParrondo, JordanSagawa, Manzano}:
\begin{equation}\label{excess_entropy}
\dot{\Sigma} \equiv -\frac{d}{dt} D(\hat{\rho}_S(t) || \hat{\pi}_S)  = \dot{S} - \dot{\Phi} \geq 0,
\end{equation}  
where $D(\hat{\rho} || \hat{\sigma}) = \tr[\hat{\rho} (\ln\hat{\rho} - \ln \hat{\sigma})]\geq 0$ is the quantum relative entropy. The term $\dot{\Phi} =\tr[\hat{\Phi} \dot{\hat{\rho}}_S]$ 
defines the effective rate at which entropy is transferred from the surroundings into the system throughout the non-equilibrium potential, 
$\hat{\Phi} = - \ln \hat{\pi}_S$, originally introduced in a classical context \cite{HatanoSasa, Prost}. The positivity of $\dot{\Sigma}$ is always guaranteed 
for quantum dynamical semigroups \cite{Spohn}, while the emerging second-law inequality in Eq.~(\ref{excess_entropy}) has been recently derived as a corollary 
from a general fluctuation theorem for a large class of quantum Completely Positive and Trace Preserving (CPTP) maps \cite{Manzano}. The effective entropy flow $\dot{\Phi}$ becomes zero for unital 
maps and reproduces the heat flow divided by temperature in the case of thermalization or Gibbs-preserving maps. 
Remarkably, in our case it can further be shown that it equals the rate at which entropy decreases in the reservoir along with relaxation (see Appendix \ref{app:reservoirentropy}).
Using the steady state $\hat{\pi}_S$ in Eq.~(\ref{steady-state}),
\begin{equation}
\dot{\Phi} = \beta~ \tr[\hat{\mathcal{S}} \hat{H}_S \hat{\mathcal{S}}^\dagger \dot{\hat{\rho}}_S] = \beta \left( \cosh(2 r) \dot{Q} - \sinh(2 r) \dot{{\cal A}} \right),
\label{entropy_flux}
\end{equation}
where we identified the heat flux entering the system from the reservoir, $\dot{Q} = \tr[\hat{H}_S \dot{\hat{\rho}}_S]$, and obtained the extra non-thermal contribution
\begin{equation}\label{extra-entropy}
\dot{{\cal A}} = \tr[\hat{{\cal A}} \dot{\hat{\rho}}_S] = -\frac{\hbar \omega}{2} \tr[(\hat{a}^{\dagger 2} e^{i \theta} + \hat{a}^2 e^{-i \theta}) \dot{\hat{\rho}}_S].
\end{equation}
Rewriting $\hat{{\cal A}} = (\hbar \omega/2) (\hat{p}_{\theta/2}^2 - \hat{x}_{\theta/2}^2)$, we see that it measures the asymmetry in the second-order 
moments of the mode quadratures, which includes both the relative shape of the variances and the relative displacements in optical phase space, being positive for $\hat{\pi}_S$. From the 
LME (\ref{master-eq-fieldmode}) we obtain that $\dot{{\cal A}}(t) = - \gamma(
{\cal A}(t) - \langle \hat{\cal A}\rangle_{\hat{\pi}_S})$, where the expected value of $\hat{\cal A}$  in the stationary state reads $\langle \hat{\cal A}\rangle_{\hat{\pi}_S} = \hbar \omega \sinh(2r) (n_{\rm th} + 1/2)$. 
Therefore, the evolution of ${\cal A}(t)$ is rather simple:  
it increases (decreases) exponentially when the interaction with the reservoir induces (reduces the) asymmetry in the phase-selected quadratures. 
As an illustrative example consider an initial state with ${\cal A} = 0$, but with diagonal elements in the $\hat{H}_S$ basis as those in $\hat{\pi}_S$. Clearly, during its relaxation $\dot{{\cal A}}>0$, 
while $\dot{Q}=0$ (see details in Appendix \ref{app:equations}), the uncertainty in $\hat{x}_{\theta/2}$ being reduced with respect to the one in $\hat{p}_{\theta/2}$ at constant energy until the steady state is reached. 
In this case, according to \eqref{entropy_flux}, $\Delta \Phi <0$, meaning that entropy is transferred from the system to the reservoir, indeed overcoming the entropy produced in the process, $\Sigma > 0$, which corresponds 
a net reduction in the system local entropy $\Delta S = \Sigma + \Delta \Phi < 0$.
The generalization of the second law [Eqs.~(\ref{excess_entropy}), (\ref{entropy_flux}), and (\ref{extra-entropy})], together with its interpretation, is our first main result.

\section{Extracting work from a single reservoir} \label{sec:single}

\begin{figure*}[t]
\includegraphics[scale=0.6]{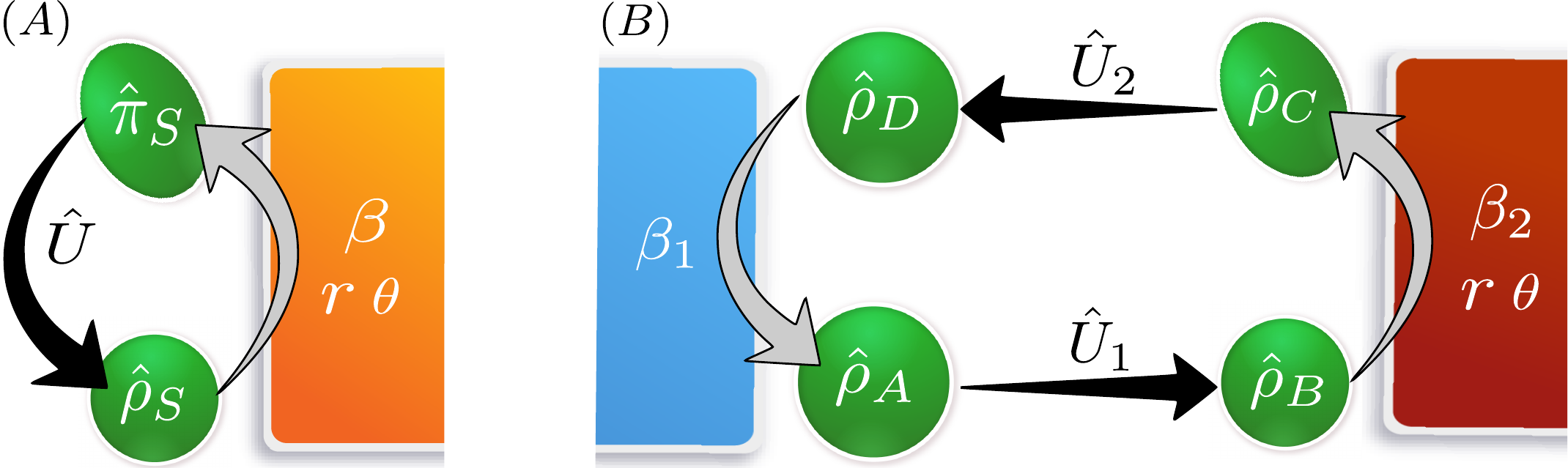}
\caption{Schematic diagrams of (A) the two-step protocol introduced to extract work from a single squeezed 
reservoir and (B) the four-step Otto-like cycle operating between reservoirs at different temperatures. The unitary $\hat{U}_1$ 
represents the adiabatic frequency modulation from $\omega_1$ to $\omega_2$, while $\hat{U}_2$ represents the convolution of the 
unitary unsqueezing the bosonic mode, $\hat{\mathcal{S}}^\dagger$, followed by adiabatic modulation from $\omega_2$ 
to $\omega_1$.} 
\label{Fig1}
\end{figure*}

As a first consequence of reservoir squeezing, we point out the possibility of cyclic work extraction from a single reservoir. 
This operation is forbidden by the second law of thermodynamics in the thermal reservoir case.
Nevertheless it becomes possible when including extra sources of coherence \cite{Photocarnot}, neg-entropy \cite{Neg-entropy}, 
or additional information reservoirs \cite{Mandal, DeffnerIR}.  
We consider a two-stroke cyclic process operated as sketched in Fig.~\ref{Fig1}(A). In the first step we start with the state $\hat{\pi}_S$ in Eq.~(\ref{steady-state}), 
and Hamiltonian $\hat{H}_S = \hbar \omega \hat{a}^\dagger \hat{a}$, implementing a unitary (isentropic) evolution $\hat{U}$, which drives the system detached from the reservoir (e.g., by 
modulating the frequency $\omega(t)$, as explained in Appendix \ref{app:experimental}). The bosonic mode ends up in some state $\hat{\rho}_S = \hat{U} \hat{\pi}_S \hat{U}^\dagger$ with 
the same Hamiltonian $\hat{H}_S$.
In this process work can be extracted by the external driving, $W_{\rm out} = \tr[\hat{H}_S \hat{\pi}_S] - \tr[\hat{H}_S \hat{\rho}_S]$, while no heat is produced. 
In the second step the system is put in contact with the squeezed thermal reservoir until it relaxes back to $\hat{\pi}_S$. 
This produces a heat flow entering from the reservoir, which equals the work extracted in the first step, $Q = \tr[\hat{H}_S \hat{\pi}_S] - \tr[\hat{H}_S \hat{\rho}_S] 
= W_{\rm out}$, as required from energy conservation. The second law Eq.~\eqref{excess_entropy}, integrated over a whole cycle, yields $- \Delta \Phi \geq 0$. Using  Eq.~(\ref{entropy_flux}), we find
\begin{equation}\label{in1}
W_{\rm out} \leq \tanh(2r) \Delta {\cal A}, 
\end{equation}
where $\Delta {\cal A} = \langle {\cal A}\rangle_{\hat{\pi}_S} - \langle {\cal A}\rangle_{\hat{\rho}_S}$.
Hence positive work may be extracted from the reservoir whenever $\Delta {\cal A} > 0$, e.g. by having $\hat{\rho}_S$ less squeezed than $\hat{\pi}_S$. Maximum work is extracted 
by requiring $\hat{\rho}_S =e^{-\beta \hat{H}_S}/Z$ (which means that $\hat{U} = \hat{\mathcal{S}}^\dagger$), as it minimizes the mean energy for a fixed entropy. In that particular case:
\begin{equation}\label{maxW}
W_{\rm max} = \hbar \omega (2 n_{\rm th} + 1) \sinh^2(r) \geq 0,
\end{equation}
which vanishes in the thermal case, $r=0$, as expected. It is worth mentioning that this process does not saturate inequality (\ref{in1}), meaning that it is not reversible, 
but an amount $\Sigma = \beta W_{\rm max}$ of entropy is produced in each cycle. Indeed reversibility conditions $(\Sigma = 0)$ can only be achieved, following Eq.~(\ref{excess_entropy}), 
in the trivial case $\hat{\rho}_S = \hat{\pi}_S$, implying $W_{\rm out} = \Delta {\cal A} = 0$.

\section{Heat engine with a squeezed thermal reservoir} \label{sec:engine}

\subsection{Optimal Otto Cycle} \label{sec:engineA}
As a second application of interest we consider a quantum heat engine 
operating between two reservoirs: a cold thermal bath at inverse temperature $\beta_1$ and a hot squeezed thermal reservoir at $\beta_2\leq\beta_1$ with squeezing parameters $\{r,\theta\}$.
The bosonic mode performs a thermodynamic four-stroke cycle [Fig.~\ref{Fig1}(B)]
as in traditional quantum Otto cycles \cite{Kieu, Rezek, Quan}, while the isentropic expansion is allowed to unsqueeze 
the mode, which in turn will allow us to exploit the full power of the squeezed thermal reservoir.

We start with our system in point A, in equilibrium with the cold thermal reservoir,   
$\hat{\rho}_A = \exp(-\beta_1 \hat{H}_1)/Z_A$, $Z_A = \tr[e^{-\beta_1 \hat{H}_1}]$.
The initial Hamiltonian is $\hat{H}_1 = \hbar \omega_1 \hat{a}^\dagger_1 \hat{a}_1$. During the first step the system 
is isolated from the reservoirs, and its frequency adiabatically modulated from $\omega_1$ to 
$\omega_2 \geq \omega_1$, without changing the populations of the energy eigenstates. 
The density matrix at point $B$ is $\hat{\rho}_B = \hat{U}_{1} \hat{\rho}_A \hat{U}_{1}^\dagger = \exp(-\beta_1 \frac{\omega_1}{\omega_2} \hat{H}_2)/Z_B$,
where $\hat{U}_1$ represents the adiabatic modulation, $Z_B = Z_A$, and the Hamiltonian is changed to $\hat{H}_2 = \hbar \omega_2 \hat{a}^\dagger_2\hat{a}_2$
during the process. The work extracted during this isentropic compression is negative (external work is needed to perform it), 
and reads $W_{AB} = \tr[\hat{H}_1 \hat{\rho}_A] - \tr[\hat{H}_2 \hat{\rho}_B] = - \hbar (\omega_2 - \omega_1) n_{\rm th}^{(1)}$, where $n_{\rm th}^{(1)} = (e^{\beta_1 \hbar \omega_1} - 1)^{-1}$. 
The Gibbs form of the state $\hat{\rho}_B$ minimizes the work lost in the compression and, as long as the system is isolated, no heat is produced in this step. 
In the second stroke, the bosonic mode is put in contact with the squeezed thermal reservoir while the frequency stays constant, resulting in 
an isochoric process where the mode relaxes to the steady-state $\hat{\rho}_C = \hat{\mathcal{S}} \exp(- \beta_2 \hat{H}_2)/Z_C \hat{\mathcal{S}}^\dagger$. 
The heat entering the system from the squeezed thermal bath in the relaxation is $Q_{BC} = \tr[\hat{H}_2 \hat{\rho}_C] - \tr[\hat{H}_2 \hat{\rho}_B]
=\hbar \omega_2(n_{\rm th}^{(2)} \cosh(2r) + \sinh^2(r) - n_{\rm th}^{(1)})$, with $n_{\rm th}^{(2)} = (e^{\beta_2 \hbar \omega_2} - 1)^{-1}$, and 
from Eq.~(\ref{extra-entropy}), we have $\Delta {\cal A}_{BC} = \hbar \omega_2 \sinh(2r)(n_{\rm th}^{(2)} + 1/2)$.

In the third stroke, the bosonic mode is again detached from the reservoirs, we apply the unitary unsqueezing to the mode, $\hat{\mathcal{S}}^\dagger$, 
and then we change its frequency adiabatically back to $\omega_1$ \cite{Niedenzu}. This process can alternatively be done by a unique tailored modulation $\omega(t)$ \cite{Galve}.
The system state at point D is then $\hat{\rho}_D = \hat{U}_{2} \hat{\rho}_C \hat{U}_{2}^\dagger = \exp(-\beta_2 \frac{\omega_2}{\omega_1} \hat{H}_1)/Z_D$, where 
$\hat{U}_2$ represents the two operations, and $Z_D = Z_C$. Consequently, the work extracted in this isentropic expansion reads 
$W_{CD} = \tr[\hat{H}_2 \hat{\rho}_C] - \tr[\hat{H}_1 \hat{\rho}_D] = \hbar \omega_2(n_{\rm th}^{(2)} \cosh(2r) + \sinh^2(r)) - \hbar \omega_1  n_{\rm th}^{(2)}$. Notice that the state $\hat{\rho}_D$ has been chosen to 
maximize the work extracted, as indicated by our previous example and Eq.~\eqref{maxW}. The cycle is closed by putting the bosonic mode in contact with the cold thermal reservoir,  
and hence relaxing back to $\hat{\rho}_A$ without varying its frequency. During the last isochoric process, the heat transferred from the cold reservoir to the system is 
$Q_{DA} = \tr[\hat{H}_1 \hat{\rho}_A] - \tr[\hat{H}_1 \hat{\rho}_D] = \hbar \omega_1 (n_{\rm th}^{(1)} - n_{\rm th}^{(2)})$. The total work extracted in the cycle is given by the contributions 
of the two isentropic strokes: 
\begin{eqnarray}\label{total-work}
 W_{\rm out} &\equiv& W_{AB} + W_{CD} = \hbar (\omega_2 - \omega_1) (n_{\rm th}^{(2)} - n_{\rm th}^{(1)}) + \nonumber \\ 
&+& \hbar \omega_2 (2 n_{\rm th}^{(2)} +1)\sinh^2(r),
\end{eqnarray}
which is nothing but the sum of the work extractable from an ideal quantum Otto cycle between two regular thermal 
reservoirs (first term), plus the work extractable from a single squeezed thermal reservoir (last term), as given by Eq.~(\ref{maxW}). 
Notice that $W_{\rm out} = Q_{BC} + Q_{DA}$, as required by the first law. 

\begin{figure}[h]
\includegraphics[scale=1.0]{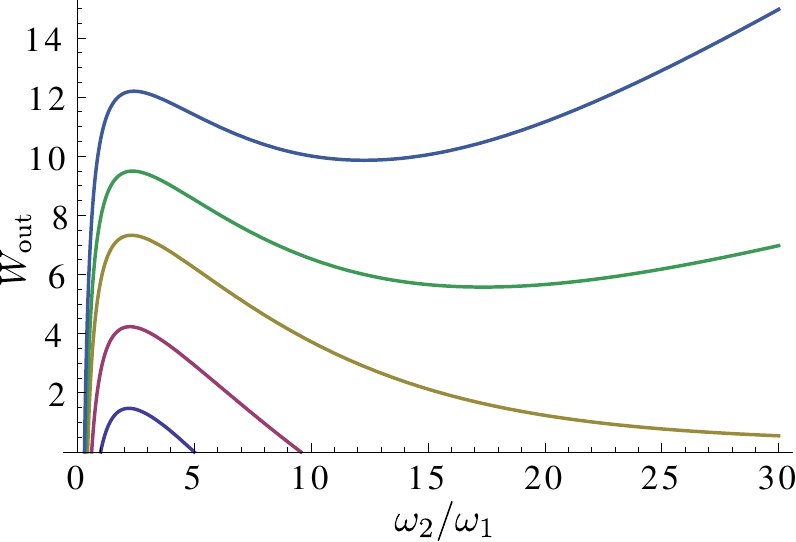}
\caption{Total work output, $W_{\rm out}$ (in units of $\hbar \omega_1$), generated in a single cycle as a function of the frequency modulation, $\omega_2/\omega_1$,
for different values of the squeezed parameter (from bottom to top) $r = (0.0, 0.5, 0.7, 0.8, 0.9)$. We used $\beta_1 = (\hbar \omega_1)^{-1}$ and $\beta_2 = 0.2 (\hbar \omega_1)^{-1}$.} 
\label{Fig2}
\end{figure}

In Fig.~\ref{Fig2} we plot the  work output of the cycle as a function of the frequency modulation $\omega_2$ (in units of $\omega_1$) for different values of the squeezed parameter. 
As we can see in the plot, the maximum power with respect to $\omega_2$ is no longer confined to the low-frequency modulation region if moderate values of the squeezing parameter are 
considered. This opens the possibility of increasing the power by frequency modulation. However the local maximum is placed at the same point as for the traditional cycle for the 
high-temperature regime, given by $\omega_2/\omega_1 = \sqrt{\beta_1(1 + 2\sinh^2(r))/\beta_2}$ \cite{LutzHE}.

\subsection{Regimes of operation} \label{sec:engineB}

\begin{figure}[t] 
 \includegraphics[width=8.0cm]{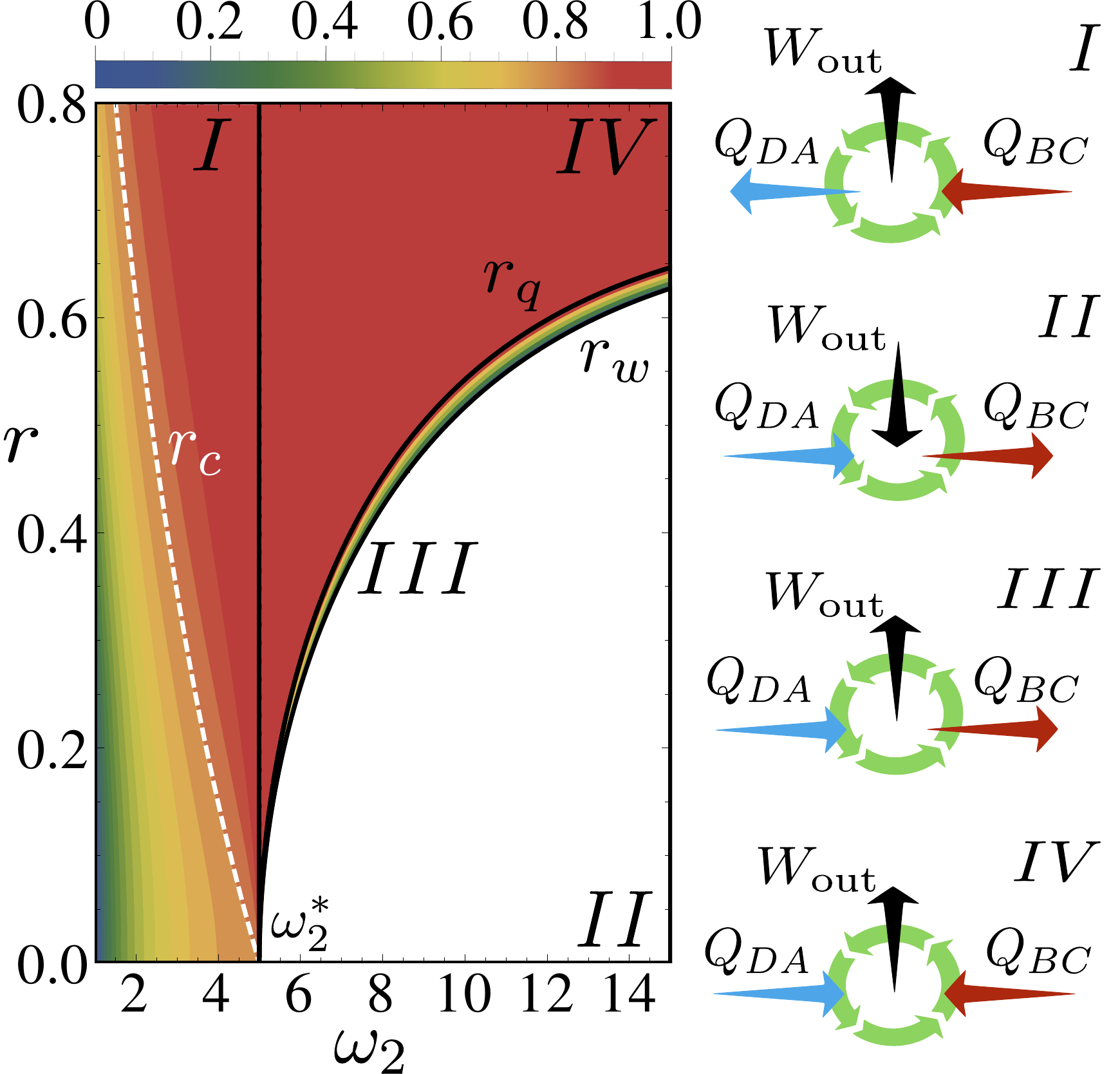}
 \caption{Phase diagram with the four regimes of operation of the cycle (I, II, III, IV) as a  function of $\omega_2$ (in units of $\omega_1$) 
 and $r$. The color scale corresponds to the energetic efficiency of the cycle $\eta = W_{\rm out}/Q_{\rm in}$ as a heat engine, for $\beta_1 = (\hbar \omega_1)^{-1}$ and 
 $\beta_2 = 0.2 (\hbar \omega_1)^{-1}$, yielding $\eta_c=0.8$. In the right side the direction of the arrows represents the sign of the energy fluxes for each regime.} 
 \label{Fig3}
\end{figure}

The above introduced cycle presents different regimes of operation depending on the squeezing $r$ and on $\omega_2$, some of them {\it forbidden} in the regular Otto cycle, which we summarize 
in the phase diagram of Fig.~\ref{Fig3}. 

{\it Region I} corresponds to a regular heat engine, for which work is extracted from the heat 
released by the hot (squeezed) reservoir, while dissipating some part in the cold thermal one. In this regime, a small frequency 
modulation, $\omega_2 \leq \omega_2^{\ast} \equiv \omega_1 \beta_{1}/\beta_{2} \Leftrightarrow n_{\rm th}^{(2)} \geq n_{\rm th}^{(1)}$, 
guarantees $W_{\rm out} \geq 0$, $Q_{BC}\geq 0$, and $Q_{DA}\leq 0$. 
The energetic efficiency, defined as the total work output, $W_{\rm out}$, divided by the input heat, $Q_{BC}$, reads:
\begin{eqnarray} \label{efficiency}
\eta = 1 - \frac{\omega_1}{\omega_2} \left( \frac{n_{\rm th}^{(2)} - n_{\rm th}^{(1)}}{(2 n_{\rm th}^{(2)} + 1)\sinh^2(r) + n_{\rm th}^{(2)} - n_{\rm th}^{(1)}} \right) ~~~~~
\end{eqnarray}
which differs from the traditional Otto cycle efficiency for adiabatic strokes, $\eta_{\rm q} = 1 - \omega_1/\omega_2$ \cite{Kieu}. 
Indeed the efficiency (\ref{efficiency}) can surpass Carnot efficiency, $\eta \geq \eta_{\rm c}= 1- \beta_2/\beta_1$, for sufficient 
large squeezing, $r \geq r_c(\omega_2)$. The Carnot line, $r_c(\omega_2)$ is depicted in Fig.~\ref{Fig3} (white dashed line) and calculated explicitly in Appendix \ref{app:otto}. Furthermore we see 
from Eq.~(\ref{efficiency}) that $\eta \rightarrow 1$ when $\omega_2^{~} \rightarrow \omega_2^{\ast}$ 
while maintaining a finite work output in the cycle,  $W_{\rm out} \rightarrow \hbar \omega_2^{\ast} (2 n_{\rm th}^{(2)} +1)\sinh^2(r)$, which is the same result as 
in the single reservoir case.

The other regions occur for large frequency modulation, $\omega_2 \geq \omega_2^\ast  \Leftrightarrow ~ n_{\rm th}^{(1)} \geq n_{\rm th}^{(2)}$, 
implying a positive amount of heat extracted from the cold reservoir, $Q_{DA} \geq 0$. 

{\it Region II}  (white area in Fig.~\ref{Fig3}) corresponds to the well-known case of a driven refrigerator: external input work is needed to pump heat from the cold to the hot reservoir $(W_{\rm out}\leq 0$ and $Q_{BC} \leq 0)$.

{\it Regions III and IV} are the most striking regimes, implying refrigeration and work extraction at the same time, 
as recently suggested in Ref.~\cite{Niedenzu}. 
From Eq.~(\ref{total-work}) one can obtain the conditions for $W_{\rm out}$ and $Q_{BC}$ to vanish, $r_w(\omega_2)$ and $r_q(\omega_2)$, respectively. Then $r \geq r_w(\omega_2)$ 
implies a positive amount of output work, whereas the heat flux entering the hot reservoir, $Q_{BC}$, is positive when  $r \geq r_q(\omega_2)$. 
We then distinguish two regions (see Fig.~\ref{Fig3}). {\it Region III} is the narrow strip between the two boundaries, 
$r_q \geq r \geq r_w$, where we obtain a refrigerator producing a positive work output while pumping heat from the cold to the hot reservoir $(W_{\rm out} \geq 0$ and $Q_{BC}\leq 0)$. 
Its efficiency as a heat engine is given by $\eta = W_{\rm out}/Q_{DA} = 1 - (\omega_2/\omega_1)( 1- \sinh^2(r)/\sinh^2(r_q))$, which varies from $0$ to $1$ between the two boundaries. 
Finally in {\it region IV} $(r \geq r_q)$, we obtain a heat engine which absorbs heat from both reservoirs, transforming it into useful work $(W_{\rm out} \geq 0$ and $Q_{BC} \geq 0)$ at 
efficiency $\eta = W_{\rm out}/Q_{\rm in} = 1$, as guaranteed by the first law. The explicit expressions for the curve $r_c$ and the boundaries $r_q$ and $r_w$ are given in Appendix \ref{app:otto}.

It is worth noticing that our results do not contradict the second law of thermodynamics, when generalized to this 
non-equilibrium situation, Eq.~\eqref{excess_entropy}. Indeed, it can be written as the positivity of the entropy production for a single cycle of the engine:
\begin{eqnarray}\label{second_law}
\Sigma_{\rm cyc}& =& -\beta_1 Q_{DA} - \beta_2 \left[\cosh(2r)Q_{BC} - \sinh(2r) \Delta {\cal A}_{BC}\right]
\nonumber \\ & \geq & 0
\end{eqnarray}
which follows from Eq.~(\ref{entropy_flux}). Using the explicit expressions of $Q_{BC}$, $Q_{DA}$, and $\Delta {\cal A}_{BC}$ for the cycle, 
we obtain that reversibility conditions $(\Sigma_{\rm cyc} = 0)$ can be only reached when $\omega_2 = \omega_2^\ast$ and $r = 0$, hence 
implying $W_{\rm out} = 0$. Finally, when the second law (\ref{second_law}) is combined with the first law, $W_{\rm out} = Q_{BC} + Q_{DA}$, 
we obtain bounds on the energetic efficiency for the heat engine regimes, $\eta \leq \eta_{\rm max}$, where:
\begin{eqnarray} \label{general_eff}
\eta_{\rm max} = \left \lbrace \begin{array}{lc}
                        1 - \frac{\beta_2}{\beta_1}\left( \cosh(2r) - \sinh(2r) \frac{\Delta {\cal A}_{BC}}{Q_{BC}} \right) & (I) \\[7pt]
			1 - \frac{\beta_1}{\beta_2\cosh(2r)} + \tanh(2r)\,\frac{\Delta {\cal A}_{BC}}{Q_{DA}} &  (III) .
                       \end{array} \right. \nonumber
\end{eqnarray}
As can be easily checked, $\eta_{\rm max} \rightarrow \eta_{\rm c}$ when $r \rightarrow 0$ in {\it region I}, while {\it regions III and IV} 
disappear in such case. The above equation is exact and generalizes previous efficiency bounds \cite{LutzHE, Abah} (only valid in the high-temperature limit) to 
any temperatures and frequencies. The expressions for $\eta_{\rm max}$ in the different operational regimes represent, together with the phase map in 
Fig. \ref{Fig3}, our second main result. The explicit formulas for $\eta_{\rm max}$ are given for the interested reader in Appendix \ref{app:otto}.

Finally, we show in Fig.~\ref{Fig4} how the efficiency $\eta$ of our cycle, even when working as a normal heat engine, Eq.~(\ref{efficiency}), can overcome the so-called 
generalized Carnot efficiency obtained in Refs.~\cite{LutzHE, Abah} by using the high-temperature approximation ($\beta_i \hbar \omega_i \ll 1$ for $i=1,2$):
\begin{equation}
 \eta_{\rm ht} = 1- \frac{\beta_2}{\beta_1 (1 + 2 \sinh^2(r))}
\end{equation}
which verifies $\eta_{\rm ht} \geq \eta_c = 1- \beta_2/\beta_1$. In contrast, our general bound, $\eta_{\rm max}\geq \eta_{\rm ht}$, obtained by applying the second law of 
thermodynamics in the full quantum regime, cannot be surpassed in any case. A complementary interpretation of the generalized second law in Eq.~({\ref{second_law}})
in terms of the free-energy released from the hot squeezed thermal reservoir, is further given in Appendix \ref{app:free-energy}.

\begin{figure}[t]
\includegraphics[scale=1.0]{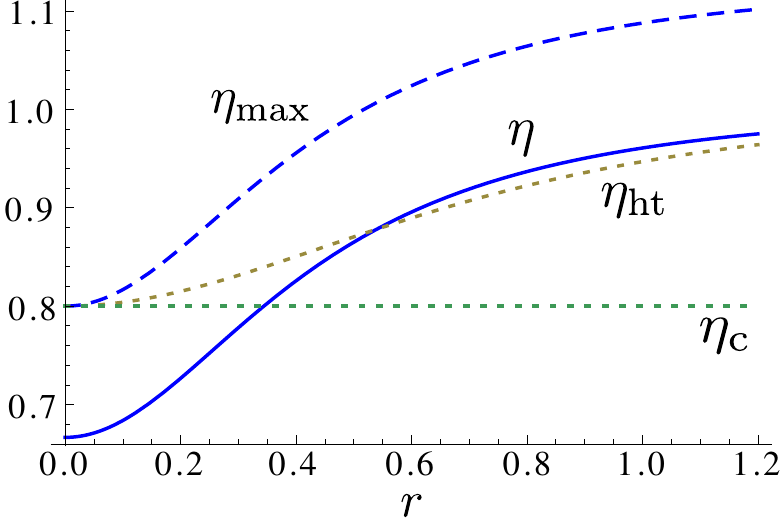}
\caption{Comparison of the efficiency of the heat engine, $\eta$, the maximum efficiency allowed by the second law, $\eta_{\rm max}$, 
the Carnot efficiency, $\eta_{c}$, and the high-temperature generalized Carnot efficiency, $\eta_{ht}$, as a function of the squeezing parameter $r$. 
The high-temperature efficiency fails to bound correctly the efficiency of the cycle for moderate values of the squeezing parameter. Here we used 
$\omega_2 = 3 \omega_1$ (i.e., $\omega_2 < \omega_2^\ast = 5 \omega_1$, corresponding to {\it region I}) and again 
$\beta_1 = (\hbar \omega_1)^{-1}$ and $\beta_2 = 0.2 (\hbar \omega_1)^{-1}$.} 
\label{Fig4}
\end{figure}

\section{Conclusions}

Squeezing constitutes a quantum thermodynamic resource from which useful work can be delivered. When squeezing is present in an otherwise 
thermal reservoir, it not only modifies the entropy flow associated with the heat exchanged with the system, but induces an extra term proportional 
to the second-order coherences, Eq.~(\ref{extra-entropy}), with a specific thermodynamic meaning. 

The non-equilibrium second law-inequality, Eq.~(\ref{excess_entropy}) with (\ref{entropy_flux}), introduces remarkable modifications 
which may give rise to novel phenomena and applications as squeezing-fueled batteries, multi task (refrigerator, heat pump, {\it and} heat 
engine) thermal machines, or a perfect heat-to-work transformer working at unit efficiency. The extra non-thermal contribution 
to the entropy transfer hints also at possible erasure devices operating below Landauer's limit \cite{Parrondo}.

In the present work, the squeezed thermal reservoir has been considered as a given thermodynamical resource. Consequently, we did not consider 
any extra energetic or thermodynamic cost associated to its creation, in the same manner as thermal reservoirs at different temperatures are considered 
as given resources for the operation of traditional heat engines. The thermodynamic cost for generating squeezing may in general depend on the specific 
configuration employed, and has been investigated e.g. in Refs. \cite{Galve, Zagoskin}.

Finally, our results may be tested as in the recent experiment of a single-ion Otto heat engine \cite{LutzHE, LutzHEexp}, with an added modification
(see Appendix \ref{app:experimental}) to additionally exploit the squeezing absorbed from the hot reservoir.

\section*{Acknowledgments}

All authors acknowledge support from COST Action MP1209. G.M. and J.M.R.P. acknowledge funding from MINECO (Grant No. FIS2014-52486-R). 
F.G. and R.Z. acknowledge funding from MINECO (Grant No. FIS2014-60343-P) and EU project QuProCS (Grant Agreement No. 641277). 
F.G. acknowledges support from 'Vicerectorat d'lnvestigaci\'o i Postgrau' of the UIB and G.M. from FPI Grant No. BES-2012-054025.

\

\appendix

\section{Collisional model}\label{app:collisional}
We construct a microscopic collisional model in order to provide a derivation of the Master Equation (\ref{master-eq-fieldmode}) in Sec. \ref{sec:thermo}, alternative to the one developed in Refs.~\cite{Walls, Scully}. 
This shall provide a more intuitive picture of the dynamical evolution generated from the squeezed thermal reservoir, while allowing examination of the thermodynamic behavior of the reservoir, 
from which we indeed benefit in the next appendices.
In the collisional model, the system bosonic mode interacts at random times, given by some rate $\mathcal{R}$, with a generic mode $k$ of the photonic environment once at a time. The 
Hamiltonian of the reservoir's mode $k$ reads $\hat{H}_R(\Omega_k) = \hbar \Omega_k \hat{b}_k^\dagger \hat{b}_k$, with $[\hat{b}_k, \hat{b}_k^\dagger]=1$. In each collision the reservoir 
mode changes, and may have a different frequency, depending on the reservoir density of states, $\varrho(\Omega_k)$, which characterizes the number of modes 
with a given frequency $\Omega_k$. For the moment let us particularize the interaction Hamiltonian to account for the interaction with a single mode in the reservoir, 
$\hat{H}_I = i \hbar g_k (\hat{a} \hat{b}_k^\dagger - \hat{a}^\dagger \hat{b}_k)$. Assuming weak coupling, such that $g_k \tau \ll 1 ~~ \forall k$, for interaction time $\tau$, the unitary 
evolution governing a single collision occurring at time $t$, reads, in the interaction frame,
\begin{eqnarray}
 \hat{U}_{I}(t + \tau, t) = T_{+} \exp \left(-\frac{i}{\hbar} \int_{t}^{t + \tau} dt_1  \hat{H}_I'(t_1) \right) ~~ \\
 ~{\rm where}~~ \hat{H}_I'(t_1) = i \hbar g_k \left( \hat{a}~\hat{b}_k^\dagger e^{-i \Delta_k  t} - \hat{a}^\dagger \hat{b}_k e^{i\Delta_k  t} \right), \nonumber
\end{eqnarray}
with $\Delta_k = \omega - \Omega_k$, and $\hat{H}_I'$ representing the interaction Hamiltonian in the interaction picture. The two-mode (total) density matrix, to second order in the 
coupling hence changes as
\begin{eqnarray}\label{rho-evolution}
& & \hat{\rho}_{\rm tot}(t + \tau, t) \simeq ~\hat{\rho}_{\rm tot}(t) - \frac{i}{\hbar} \int_t^{t + \tau} dt_1 [\hat{H}_I'(t_1), \hat{\rho}_{\rm tot}(t)]  \nonumber \\ 
& & - \frac{1}{\hbar^2} \int_t^{t + \tau} dt_2 \int_{t}^{t_2} dt_1  [\hat{H}_I'(t_2), [\hat{H}_I'(t_1),  \hat{\rho}_{\rm tot}(t)]],
\end{eqnarray}
where we obtain for the first order commutator
\begin{equation}
 [\hat{H}_I'(t_1), \hat{\rho}_{\rm tot}(t)] = i \hbar g_k \left( [\hat{a} \hat{b}_k^\dagger, \rho_{\rm tot}(t)]e^{- i \Delta_k t_1} - {\rm h.c.} \right) \nonumber
\end{equation}
and for the second-order one
\begin{widetext}
\begin{eqnarray}
[\hat{H}_I'(t_2), [\hat{H}_I'(t_1),  \hat{\rho}_{\rm tot}(t)]] = - \hbar^2 g_k^2 \left( 
[\hat{a}^\dagger \hat{b}_k ,[\hat{a}^\dagger \hat{b}_k, \rho_{\rm tot}(t)]] e^{i \Delta_k (t_1 + t_2)}  
 -  [\hat{a}~\hat{b}_k^\dagger ,[\hat{a}^\dagger \hat{b}_k, \rho_{\rm tot}(t)]] e^{i \Delta_k (t_1 - t_2)} + {\rm h.c.} \right).
\end{eqnarray}
\end{widetext}

The reduced evolution in the system and in the reservoir mode, can be obtained by  partial tracing of Eq.~(\ref{rho-evolution}) over the corresponding degrees of 
freedom. We also assume $\hat{\rho}_{\rm tot}(t) = \hat{\rho}_S(t) \otimes \hat{\rho}_R^{(k)}$; i.e., the system mode always interacts with a ``fresh'' reservoir mode $k$
in the same squeezed thermal state at inverse temperature $\beta$, and squeezing parameters $r \geq 0$ and $\theta \in [0, 2\pi]$:
\begin{eqnarray}\label{rhoR}
 \hat{\rho}_R^{(k)} = \hat{\mathcal{S}}_k \frac{e^{-\beta \hat{H}_R(\Omega_k)}}{Z_R} \hat{\mathcal{S}}_k^\dagger 
 = \sum_\nu \frac{e^{- \beta \hbar \Omega_k \nu}}{Z_R} \hat{\mathcal{S}}_k \ket{\nu_k} \bra{\nu_k} \hat{\mathcal{S}}_k^\dagger ~~~~~~
\end{eqnarray}
where $\hat{\mathcal{S}}_k \equiv \exp{\frac{r}{2}(b_k^2 e^{- i \theta} - b_k^{\dagger 2} e^{i \theta})}$, stands for the (unitary) squeezing operator on the reservoir mode $k$, 
and in the last equality we decomposed the Gibbs state in its Fock basis $\{ \ket{\nu_k}\}$. It is easy to see from the above equation that the eigenvalues and eigenvectors 
of $\hat{\rho}_R$ are given by
\begin{equation}\label{eigenbasis}
 \epsilon_\nu^{(k)} = \frac{e^{- \beta \hbar \Omega_k \nu}}{Z_R} ~~~;~~~ \ket{\epsilon_\nu^{(k)}} = \hat{\mathcal{S}}_k \ket{\nu_k},
\end{equation}
i.e., the state $\hat{\rho}_R^{(k)}$ can be viewed as a classical mixture of squeezed Fock states $\ket{\epsilon_v^{(k)}}$ with Boltzmann weights $\epsilon_\nu^{(k)}$.

The Master Equation can be constructed from the following coarse-grained derivative for the system mode. During some small interval of time $\delta t \sim \mathcal{R}^{-1}$ 
(but $\delta t \gg \tau$), for which only one interaction occurs:
\begin{equation}\label{coarse}
 \dot{\hat{\rho}}_{S}(t) \simeq  \frac{1}{\delta t}[\hat{\rho}_S(t + \delta t) -\hat{\rho}_S(t )] = \mathcal{R}[\hat{\rho}_S(t + \tau) -\hat{\rho}_S(t )]    \nonumber
\end{equation}
where $\hat{\rho}_S(t) = \tr_R[\hat{\rho}_{\rm tot}(t)]$ and the second equality follows from the fact that the density matrix in the interaction picture does not change 
when no interaction with the reservoir takes place. This is valid when the reservoir modes always have the same frequency $\Omega_k$, but if we 
want to take into account that the reservoir contains many frequencies, the above equation should be averaged over the the reservoir density of states:
\begin{equation}\label{coarse}
 \dot{\hat{\rho}}_{S}(t) \simeq  \mathcal{R} \sum_k \varrho(\Omega_k)[\hat{\rho}_S(t + \tau) -\hat{\rho}_S(t )].   
\end{equation}

Now performing the time integrals, the partial trace, and substituting the reservoir expectation values: 
\begin{eqnarray}\label{av}
\langle b_k \rangle_{\hat{\rho}_R^{(k)}} &=& 0 ~~~~~~~~~~~~~~~~  \langle b_k^\dagger \rangle_{\hat{\rho}_R^{(k)}} = 0 ~~ \nonumber \\ 
\langle b_k^2 \rangle_{\hat{\rho}_R^{(k)}} &=& M(\Omega_k) ~~~~~~ \langle b_k^{\dagger 2} \rangle_{\hat{\rho}_R^{(k)}} =M^\ast(\Omega_k) \\
\langle b_k^\dagger b_k \rangle_{\hat{\rho}_R^{(k)}} &=& N(\Omega_k) ~~~~~~ \langle b_k b_k^\dagger \rangle_{\hat{\rho}_R^{(k)}} = N(\Omega_k) + 1 ~~\nonumber
\end{eqnarray}
where $N(\Omega_k) = n_{\rm th}(\Omega_k) \cosh(2r) + \sinh^2(r)$ and $M(\Omega_k) =-\sinh(r)\cosh(r) (2 n_{\rm th}(\Omega_k) + 1) e^{i \theta}$ with 
$n_{\rm th}(\Omega_k) = (e^{\beta \hbar \Omega_k}-1)^{-1}$, one arrives at
\begin{eqnarray}\label{MEapp}
 \dot{\hat{\rho}}_S(t) &=& - i[\Delta \hat{H}_S,  \hat{\rho}_S(t)] +  \\
 & +& \Gamma_e  \left( \hat{a} \hat{\rho}_S(t) \hat{a}^\dagger - \frac{1}{2}\{\hat{a}^\dagger \hat{a} , \hat{\rho}_S(t) \} \right)  \nonumber \\ 
 & +& \Gamma_a   \left( \hat{a}^\dagger \hat{\rho}_S(t) \hat{a} - \frac{1}{2}\{\hat{a} \hat{a}^\dagger , \hat{\rho}_S(t) \} \right)  \nonumber \\ 
 & -& \Gamma_{s}  \left( \hat{a}^\dagger \hat{\rho}_S(t) \hat{a}^\dagger - \frac{1}{2}\{\hat{a}^{\dagger 2} , \hat{\rho}_S(t) \} \right) \nonumber \\ 
 & -& \Gamma_{s}^\ast \left( \hat{a} \hat{\rho}_S(t) \hat{a} - \frac{1}{2}\{\hat{a}^2 , \hat{\rho}_S(t) \} \right), \nonumber
\end{eqnarray}
where we identified the following decay factors charactering the time scales of emission/adsorption processes and squeezing:
\begin{eqnarray}\label{gamma1}
 \Gamma_e &\equiv& \mathcal{R} \tau^2 \int_0^\infty d \Omega J(\Omega) {\rm sinc}^2(\Delta \tau /2) (N(\Omega) + 1)  \nonumber \\
 \Gamma_a &\equiv& \mathcal{R} \tau^2 \int_0^\infty d \Omega J(\Omega) {\rm sinc}^2(\Delta \tau /2) N(\Omega) \\
 \Gamma_{s} &\equiv& \mathcal{R} \tau^2 \int_0^\infty d \Omega J(\Omega) {\rm sinc}^2(\Delta \tau /2) M(\Omega) e^{i \Delta (2t + \tau)} \nonumber
 \end{eqnarray}
together with the reservoir-induced frequency shift:
\begin{eqnarray}
&& \Delta \hat{H}_S = \mathcal{R} \int_0^\infty d\Omega J(\Omega) \frac{\tau}{\Delta} \lbrace \hat{a}^\dagger \hat{a} ({\rm sinc}(\Delta \tau/2)\cos(\Delta \tau/2) - 1)  \nonumber \\ 
&& ~~+  1 - {\rm sinc}(\Delta \tau/2)\left(2N(\Omega)(\cos(\Delta \tau/2) - 1) + e^{i\Delta \tau/2}\right) \rbrace \nonumber 
\end{eqnarray}
which will be neglected in the following. In the above equations we introduced the reservoir spectral density $J(\Omega) = \sum_k g_k^2 \varrho(\Omega_k) \delta_D(\Omega - \Omega_k)$ 
and took the continuum limit. Notice that the three integrals in Eqs.~(\ref{gamma1}) are weighed by the function ${\rm sinc}^2(\Delta \tau/2)$. As this factor is highly peaked around 
$\Delta = 0$ (this is $\Omega = \omega$), it acts as a Dirac delta function ($\delta_D(\Delta \tau /2)$) when integrating over the reservoir frequencies, meaning that the effect of 
detuned modes in the reservoir is very weak in comparison with the resonant ones \cite{Sch01}. This would imply that
\begin{eqnarray}\label{gamma2}
 \Gamma_e &\simeq&  \mathcal{R} \tau^2 J(\omega) (N(\omega) + 1) \equiv \gamma (N +1), \nonumber \\
 \Gamma_a &\simeq&  \mathcal{R} \tau^2 J(\omega) N(\omega) \equiv \gamma N,  \\
 \Gamma_s &\simeq&  \mathcal{R} \tau^2 J(\omega) M(\omega) \equiv \gamma M, \nonumber
\end{eqnarray}
and we obtain an effective decay rate $\gamma = \mathcal{R} \tau^2 J(\omega)$ characterizing the global system-reservoir interaction dynamics, proportional 
to the density of resonant modes in the reservoir. Furthermore, here and in the following we denote $N \equiv N(\omega)$ and $M \equiv M(\omega)$.

As can be easily checked the above Master Equation (\ref{MEapp}) with Eqs.~(\ref{gamma2}) is fully equivalent to the Master Equation (\ref{master-eq-fieldmode}), by 
identifying the Lindblad operators 
\begin{equation}
\hat{R}_{-} = \sqrt{\gamma (n_{\rm th}(\omega) + 1)}\hat{R}, ~~~~ \hat{R}_{+} = \sqrt{\gamma n_{\rm th}(\omega)}\hat{R}^\dagger,
\end{equation}
with $\hat{R} = \hat{a} \cosh(r) + \hat{a}^\dagger \sinh(r)e^{i \theta}$. The consistency of the present derivation is ensured by the separation of the time scales, 
$\gamma \ll \mathcal{R} \ll \tau^{-1}$, which are analogous to the approximations usually employed in the derivation of the perturbative dynamics of the celebrated one-atom maser.

\section{Reservoir entropy changes}\label{app:reservoirentropy}

In the main text we claim that the effective entropy flow, $\dot{\Phi}$, appearing in the generalized second law inequality, Eq.~(\ref{entropy_flux}) in Sec. \ref{sec:thermo}, equals 
the entropy decrease in the reservoir due to the interaction with the bosonic mode. We demonstrate here this relation from the collisional model introduced 
above. Indeed we can estimate the reservoir entropy change during the evolution by constructing, 
analogously to what have been done for the system bosonic mode, a coarse-grained time derivative,
\begin{equation}
\dot{\hat{\rho}}_{R}^{(k)} \simeq  \frac{1}{\delta t}[\hat{\rho}_R^{(k)}(t + \delta t) -\hat{\rho}_R^{(k)}] = \mathcal{R} [\hat{\rho}_R^{(k)}(t + \tau) -\hat{\rho}_R^{(k)}]  \nonumber \\
\end{equation}
for the interaction between the system and a particular mode $k$ in the reservoir. We obtain
\begin{eqnarray} \label{MEres}
\dot{\hat{\rho}}_{R}^{(k)} &=& - i [\Delta \hat{H}_R(\Omega_k) , \hat{\rho}_R^{(k)}] 
+  [\epsilon_k^\ast \langle \hat{a}\rangle_t \hat{b}_k^\dagger - \epsilon_k \langle \hat{a}^\dagger \rangle_t \hat{b}_k , \hat{\rho}_R^{(k)}] + \nonumber \\
 &+& c_k \langle \hat{a} \hat{a}^\dagger \rangle_{t} \left(\hat{b}_k \hat{\rho}_R^{(k)} \hat{b}_k^\dagger - \frac{1}{2}\{ \hat{b}_k^\dagger \hat{b}_k, \hat{\rho}_R^{(k)}\} \right)  \nonumber \\
 &+& c_k \langle \hat{a}^\dagger \hat{a} \rangle_{t} \left(\hat{b}_k^\dagger \hat{\rho}_R^{(k)} \hat{b}_k - \frac{1}{2}\{ \hat{b}_k \hat{b}_k^\dagger, \hat{\rho}_R^{(k)}\} \right) \nonumber \\
 &-& c_k e^{-i \Delta_k (2t + \tau)} \langle \hat{a}^2 \rangle_{t} \left( \hat{b}_k^\dagger \hat{\rho}_R^{(k)} \hat{b}_k^\dagger - \frac{1}{2}\{\hat{b}_k^{\dagger 2} , \hat{\rho}_R^{(k)} \} \right) \nonumber \\
 &-& c_k e^{i \Delta_k (2t + \tau)} \langle \hat{a}^{\dagger 2} \rangle_{t} \left( \hat{b}_k \hat{\rho}_R^{(k)} \hat{b}_k - \frac{1}{2}\{\hat{b}_k^2 , \hat{\rho}_R^{(k)} \} \right) 
 \end{eqnarray}
where $\langle \cdot \rangle_t = \tr_S[(\cdot) \hat{\rho}_S(t)]$. We defined
\begin{eqnarray}
 \epsilon_k &\equiv& \mathcal{R}~ \tau g_k ~{\rm sinc}(\Delta_k \tau/2) e^{i\Delta_k (t + \tau/2)}, \nonumber \\
 c_k &\equiv& \mathcal{R}~ \tau^2 g_k^2 ~{\rm sinc}^2(\Delta_k \tau /2) 
 \end{eqnarray}
together with the mode dependent frequency-shift in the reservoir,
\begin{eqnarray}
 & & \Delta \hat{H}_R(\Omega_k) \equiv \mathcal{R} \frac{g_k^2 \tau}{\Delta_k} \lbrace \hat{b}^\dagger \hat{b} ({\rm sinc}(\Delta_k \tau/2)\cos(\Delta_k \tau/2) - 1) + \nonumber \\
 && + 1 - {\rm sinc}(\Delta_k \tau/2)\left(2\langle \hat{a}^\dagger \hat{a} \rangle_t (\cos(\Delta_k \tau/2) - 1) + e^{i\Delta_k \tau/2}\right) \nonumber \}
\end{eqnarray}
which is analogous to the system frequency shift, and will be neglected as well. Notice that Eq.~(\ref{MEres}) gives us the average evolution of the reservoir mode $k$ when they interact 
once at a time with the system at random times (as specified by the rate $\mathcal{R}$). However, we do not know the frequency of the reservoir mode interacting with the system in each 
collision, so we must assume that the system interacts with all modes in the reservoir with certain probability. Therefore the average reservoir entropy change due to the entropy 
change in all reservoir modes during the evolution should read
\begin{equation}
 \dot{S}_R = \sum_k \varrho(\Omega_k) \dot{S}_R^{(k)} = - \sum_k \varrho(\Omega_k) \tr_R[\dot{\hat{\rho}}_R^{(k)} \ln \hat{\rho}_R^{(k)}]. \nonumber
\end{equation}
In the following we introduce the explicit form of $\hat{\rho}_R^{(k)}$ as given in Eq.~(\ref{rhoR}) into the above expression for the average reservoir entropy change, and exploit 
Eq.~(\ref{MEres}). We obtain
\begin{eqnarray}
 \dot{S}_R &=& \beta \sum_k \varrho(\Omega_k)~\tr_R[\dot{\hat{\rho}}_R^{(k)} ~ \hat{\mathcal{S}}_k \hat{H}_R(\Omega_k) \hat{\mathcal{S}}^\dagger_{k}] = \nonumber \\ 
 &=& - \beta ~\tr_S[\dot{\hat{\rho}}_S(t) \hat{\mathcal{S}} \hat{H}_S \hat{\mathcal{S}}^\dagger] = - \dot{\Phi},
\end{eqnarray}
where the second line follows after a little operator algebra, by expanding $\hat{\mathcal{S}}_k \hat{H}_R(\Omega_k) \hat{\mathcal{S}}^\dagger_{k}$ and using Eqs.~(\ref{MEres}) and (\ref{MEapp}). 
As a hint, first notice that the first-order term in Eq.~(\ref{MEres}) does not contribute to the entropy. Second, notice that once the trace over the reservoir degrees of freedom has been 
performed, one can take the continuum limit over the reservoir spectra by introducing the spectral density, $J(\Omega)$, to recover the system ME decay factors in Eq.~(\ref{gamma2}) after integrating 
over frequencies.

Henceforth the entropy flow entering the system during the evolution, as given by $\dot{\Phi}(t) = - \tr[\dot{\hat{\rho}}_S(t) \ln \hat{\pi}_S]$, Eq.~(\ref{entropy_flux}) in Sec. \ref{sec:thermo}, 
is the average entropy lost in the reservoir in the sequence of collisions. This implies that the excess (or non-adiabatic) entropy production 
\cite{EspositoTF,JordanParrondo,JordanSagawa, Manzano}, $\Sigma$ in Eq.~(\ref{excess_entropy}), corresponds indeed to the total entropy produced in the process. In terms of the rates, 
\begin{equation}
 \dot{\Sigma} \equiv - \frac{d}{dt} D(\hat{\rho}_S(t) || \hat{\pi}_S) = \dot{S} + \dot{S}_R \geq 0,
\end{equation}
where $D(\hat{\rho}||\sigma) = \tr[\hat{\rho}(\ln \hat{\rho} - \ln \hat{\sigma})]$ is the quantum relative entropy. As a consequence, the house-keeping (or adiabatic) 
contribution due to non-equilibrium external constraints \cite{EspositoTF, JordanParrondo} is always zero in the present case. An important consequence of the 
above finding is that no entropy is produced in order to maintain the non-equilibrium steady state $\hat{\pi}_S$, Eq.~(\ref{steady-state}), provided we have 
access to an arbitrarily big ensemble of reservoir modes in the state $\hat{\rho}_R$.

\section{Equations of motion}\label{app:equations}
From the Master Equation (\ref{master-eq-fieldmode}) in Sec. \ref{sec:thermo}, one can derive the following equations of motion for the expectation values of the Lindblad 
operators and its combinations:
\begin{eqnarray}
 \frac{d}{dt} \langle \hat{R} \rangle_t = - \frac{\gamma}{2} \langle \hat{R} \rangle_t ~~,~~ \frac{d}{dt} \langle \hat{R}^2 \rangle_t = - \gamma \langle \hat{R}^2 \rangle_t \nonumber \\
 \frac{d}{dt} \langle \hat{R}^\dagger \hat{R} \rangle_t = - \gamma \left( \langle \hat{R}^\dagger \hat{R} \rangle_t - n_{\rm th}(\omega) \right),
\end{eqnarray}
where again we denoted $\langle \cdot \rangle_t = \tr_S[(\cdot) \hat{\rho}_S(t)]$. They can then be employed to explicitly asses the dynamics of the different contributions appearing in the 
effective entropy flow, $\dot{\Phi}$ in Eq.~(\ref{entropy_flux}). Indeed, by rewriting
\begin{eqnarray}
 \hat{a} = \hat{R} \cosh(r) - \hat{R}^\dagger \sinh(r)e^{i \theta} 
 \end{eqnarray}
and substituting it into the expressions $\dot{Q}(t) = \dot{U}_S(t) = \tr[\hat{H}_S \dot{\hat{\rho}}_S(t)]$ for the heat flux entering from the reservoir, and 
$\dot{\mathcal{A}}(t) = \tr[\hat{\mathcal{A}} \dot{\hat{\rho}}_S(t)]$ with $\hat{\mathcal{A}} = -\frac{\hbar \omega}{2} (\hat{a}^{\dagger 2} e^{i \theta} +  
\hat{a}^{2} e^{-i \theta})$, for the extra non-thermal contribution we obtain
\begin{eqnarray}
 \dot{Q}(t) &=& - \gamma \left( U_S(t) - \langle \hat{H}_S \rangle_{\hat{\pi}_S} \right) \nonumber \\
 \dot{\mathcal{A}}(t) &=& - \gamma \left( \mathcal{A}(t) - \langle \hat{\mathcal{A}} \rangle_{\hat{\pi}_S} \right).
\end{eqnarray}
In the above equations we introduced the steady state values $\langle \hat{H}_S \rangle_{\hat{\pi}_S} = \hbar \omega N$ and $\langle \hat{\mathcal{A}} \rangle_{\hat{\pi}_S} = \hbar \omega |M|$, 
where $\hat{\pi}_S$ is given in Eq.~(\ref{steady-state}), and the quantities $N$ and $M$ are defined in (\ref{av}) for the resonance frequency $\omega$. We notice that both flows behave monotonically, 
yielding an exponential decay.

\section{Optimal Otto cycle details}\label{app:otto}

Quantum Otto heat engines are characterized by the implementation on the working fluid of a four-stroke cycle in which isentropic and isochoric 
processes are alternated. In the case of a bosonic mode, the isentropic (unitary) strokes are implemented by means of external modulation of 
the mode frequency, while isochoric ones are obtained by letting the frequency remain constant, while relaxing in contact with thermal reservoirs at different 
temperatures. In such case adiabatic modulation of the frequency leads to both maximum work extraction and highest efficiencies. This fact can be understood 
from a simple argument: as long as the mode state before the isentropic stroke, say $\hat{\rho}_{\rm i}$, is fixed by the previous thermalization step, we have that 
the work extracted in the process, $W_{\rm stroke} = \tr[\hat{H}_{\rm i} \hat{\rho}_{\rm i}] - \tr[\hat{H}_{\rm f} \hat{\rho}_{\rm f}]$, is minimized when $\hat{\rho}_{\rm f}$ 
(the state after modulation) has minimum energy for a fixed entropy. This occurs, of course, when it has Gibbs form $\hat{\rho}_{\rm f} = \exp(-\beta \hat{H}_{\rm f})/Z_{\rm f}$ 
for some $\beta$, which is the case when the modulation is implemented adiabatically. Moreover the quantum friction in such case is zero, as the non diagonal 
elements of the mode state in its instantaneous Hamiltonian basis, are zero during the whole cycle.

However, in the case in which squeezed thermal reservoirs are considered, the above situation is slightly modified. In Sec.~\ref{sec:engineA}, we introduced a modification 
in the traditional Otto cycle which maximizes the work extracted by applying the above argument to the new situation. In contrast to Refs.~\cite{LutzHE,Abah}, 
we require an isentropic stroke driving the state after relaxation in the presence of the squeezed thermal reservoir, $\hat{\rho}_C$, to a perfect 
Gibbs state with respect to the final Hamiltonian at the end of the stroke ($\hat{\rho}_D$). This operation can be achieved by first unsqueezing 
the mode and then applying regular adiabatic modulation, or by an unique tailored modulation \cite{Galve} (see Appendix \ref{app:experimental}). As a consequence, the power 
output defined as the work extracted in a single cycle, Eq.~(\ref{total-work}), divided by its duration is maximized. 

This way of performing the cycle is the key to obtaining the {\it forbidden} regimes of operation we report in Sec. \ref{sec:engineB}, illustrated in the phase diagram of Fig.~\ref{Fig3}. 
Here we give the explicit expressions obtained for the boundaries delimiting the operational {\it regions I, II, III, and IV}. The quantities $r_q$ and $r_w$ are 
defined via
\begin{eqnarray}
 \sinh^2(r_q) &=& (n_{\rm th}^{(1)} -  n_{\rm th}^{(2)})/(2  n_{\rm th}^{(2)} + 1) \nonumber \\
 \sinh^2(r_w) &=& (1 - \omega_1/\omega_2) \sinh^2(r_q)  
\end{eqnarray}
while  $\omega_2^\ast = \omega_1 \beta_1 / \beta_2$ (black solid lines in Fig.~\ref{Fig3}). We remember that $n_{\rm th}^{(1)}= (e^{\beta_1 \hbar \omega_1} - 1)^{-1}$ and 
$n_{\rm th}^{(2)}= (e^{\beta_2 \hbar \omega_2} - 1)^{-1}$. In the other hand, the amount of squeezing needed to overcome Carnot's efficiency in {\it region I} is given by: 
\begin{equation}
 \sinh^2(r_c) = \left(\omega_2^\ast/ \omega_2 - 1 \right) (n_{\rm th}^{(2)} - n_{\rm th}^{(1)})/(2 n_{\rm th}^{(2)} + 1)
\end{equation}
(white dashed line in Fig.~\ref{Fig3}). Notice that $r_c$ is only well defined in {\it region I}, for $\omega_2 \leq \omega_2^\ast$, implying 
$n_{\rm th}^{(2)} \geq n_{\rm th}^{(1)}$ and hence heat dissipation in the cold thermal reservoir, while $r_q$ and $r_w$ are well defined for $\omega_2 \geq \omega_2^\ast$ which ensures 
$n_{\rm th}^{(1)} \geq n_{\rm th}^{(2)}$ and hence refrigeration of the cold reservoir. It is also worth noticing from the above equations that while the different regions in 
Fig.~\ref{Fig3} may be scaled depending on the temperatures of the reservoirs, they always have the same shape.

Finally, we give for the interested reader the explicit expressions of the efficiency bound, $\eta_{\rm max}$ of Sec. \ref{sec:engineB}. For our cycle operating in the regime 
$\omega_2 \leq \omega_2^\ast$ ({\it region I}),
\begin{equation}
 \eta_{\rm max}^{(I)} = 1 - \frac{\beta_2}{\beta_1}\frac{(2 n_{\rm th}^{(2)} +1) - \cosh(2r) ( 2 n_{\rm th}^{(1)} +1 ) }{\cosh(2r) ( 2 n_{\rm th}^{(2)} +1 )- ( 2 n_{\rm th}^{(1)} +1)} 
\end{equation}
which collapses to Carnot efficiency when $r \rightarrow 0$. On the other hand, for {\it region III} we obtain
\begin{equation}
 \eta_{\rm max}^{(III)} = 1 - \frac{\beta_1}{\beta_2 \cosh(2r)} + \frac{\omega_2}{\omega_1} \frac{\tanh(2r) \sinh(2r)}{2\sinh^2(r_q)}, 
\end{equation}
only valid when $\omega_2 \geq \omega_2^\ast$ and $r_w \leq r \leq r_q$. Finally we remember that in {\it region IV} we have 
$\eta_{\rm max}^{(IV)} = \eta = W_{\rm out}/(Q_{BC} +Q_{DA}) = 1$, which follows from energy conservation.

\section{Squeezing as a source of free-energy}\label{app:free-energy}
Here we provide an interpretation of the squeezed thermal reservoir as a free-energy source, which enables work extraction 
in the quantum Otto cycle discussed in Sec. \ref{sec:engine}. The non-equilibrium free energy is a powerful concept in non-equilibrium 
thermodynamics and specifically in thermodynamics of information \cite{Parrondo}. It is defined as a property of a system 
in some arbitrary state $\hat{\rho}$ with Hamiltonian $\hat{H}$, with respect to a thermal reservoir at temperature $T$, as
\begin{equation}
 \mathcal{F}(T) = \langle \hat{H} \rangle_{\hat{\rho}} - k_B T S(\hat{\rho}),
\end{equation}
where $S(\hat{\rho})$ is the von Neumann entropy of the system state for the quantum case. The most important property of the 
non-equilibrium free-energy is that its variation measure the maximum work which can be extracted when letting the system 
equilibrate to temperature $T$ in an intelligent way \cite{Parrondo, Popescu}.    

In order to apply this concept in our situation we proceed by using the fact that the entropy transfer between system and reservoir 
equals (minus) the entropy change in the squeezed reservoir during the corresponding relaxation stroke of the Otto cycle, $\Delta \Phi_{BC} = - \Delta S_{R_2}$, 
as we showed in Appendix \ref{app:reservoirentropy}. When this point is combined with the first law in the cycle, $W_{\rm out} = Q_{DA} + Q_{BC}$, 
we can rewrite the second law inequality in Eq.~(\ref{second_law}) as
\begin{equation}
W_{\rm out} \leq \Delta \mathcal{F}_2(T_1), 
\end{equation}
where $\Delta \mathcal{F}_2(T_1) = Q_{BC} + k_B T_1 \Delta S_{R_2}$ is the loss of (non-equilibrium) free-energy in the hot 
squeezed thermal reservoir in a cycle, with respect to the cold thermal reservoir at temperature $T_1$. Furthermore this free-energy change 
can be decomposed into two separate contributions by using the explicit expression of the entropy flow, Eq.~(\ref{entropy_flux}) in Sec. \ref{sec:thermo}:
\begin{eqnarray}
 \Delta \mathcal{F}_2(T_1) &=&  \left(1 -\frac{T_1}{T_2}\right)Q_{BC} ~~+  \\ 
&+& ~ \frac{T_1}{T_2}\left( \sinh(2r) \Delta \mathcal{A}_{BC} - 2 \sinh^2(r) Q_{BC} \right). \nonumber
\end{eqnarray}
The two terms correspond respectively to the free-energy available as a consequence of the temperature gradient between two thermal reservoirs 
(first term), and the one provided by the non-equilibrium squeezing effects (second term). The first term is always positive when $Q_{BC}>0$, 
meaning that free-energy is available from the spontaneous flux of heat from a hot reservoir to a colder one. The second term, purely due to 
squeezing in the reservoir, is instead positive when squeezing is present, $r > 0$, and the following inequality is verified:
\begin{equation}\label{ineq-free}
\Delta \mathcal{A}_{BC} \geq  \tanh(r) Q_{BC}.
\end{equation}
This implies that the entropic flux of second-order coherences from the squeezed thermal reservoir [see Eq.~(\ref{extra-entropy}) in Sec. \ref{sec:thermo}], 
acts as an independent source of free-energy when the above inequality is fulfilled, increasing the work that can be extracted in the cycle. 
Furthermore it can be positive even if $Q_{BC} \leq 0$, and compensate the thermal term (which in this case would be negative), in order 
to enable work extraction, as is the case of {\it region III} of the phase diagram in Fig.~\ref{Fig3}.

\section{Experimental realization}\label{app:experimental}

We build on the single trapped-ion Otto cycle proposed in Ref.~\cite{LutzHEtheo} and successfully experimentally realized in Ref.~\cite{LutzHEexp} only recently.
There, a trapped ion in a tapered Paul trap is subjected to adiabatic frequency modulations for the isentropic strokes of the cycle.
The thermalization strokes are implemented by laser cooling with variable detuning (and thus final temperature).
The same authors proposed theoretically to enhance the cycle by having a hot bath which is squeezed \cite{LutzHE}, finding an increase of the efficiency at
maximum power. The squeezed hot reservoir was effectively implemented by having the ion thermalize (the hot reservoir does laser cooling)
and then squeezing it, resulting in a final thermal squeezed state (as if the bath were squeezed). Such squeezing operation consists of 
quenching the ion frequency from $\omega$ to $\omega+\Delta\omega$ ``for a quarter of the oscillation period'', then to $\omega-\Delta\omega$ ``for another quarter, before it is
returned to its initial value'' $\omega$ (notice that the authors \cite{LutzHE} are talking about periods of different duration, since 
the frequency of oscillations differ by $2\Delta\omega$, and this has be to be carefully accounted for in the experiment). This operation can be easily understood from Fig.~1
in Ref.~\cite{ajanzky}, by noting that suddenly increasing (decreasing) the frequency squeezes (stretches) the $x$ variance, while at constant frequency the Wigner function just rotates at
that frequency. Finally, the authors propose to output the work of the cycle (done in the radial coordinate of the ion) into the axial coordinate (the two motions are
coupled due to the tapered geometry of the trap). In this sense, the engine does work on the axial motion and the working substance is the radial motion.

In our cycle, we are adding an extra step which tries to profit from the squeezing absorbed from the hot reservoir to produce work.
In terms of operations we could just use the described proposal for the $CD$ branch (operation $\hat{U}_2$ in Sec. \ref{sec:engineA}), by just reversing 
the modulation, which would remove the squeezing from the system. In this way, though, the work would be wasted into the frequency quencher (the 
electronics of the experiment). In order to profit from the squeezing absorbed from the reservoir, we should be able to transfer it 
to some fruitful target. One possibility is to wait for the axial-radial coupling to exchange the squeezing in the radial direction (so the axial 
component absorbs all energy from the radial one). The detailed dynamics should be studied thoroughly to check for limitations, though. Another possibility,
though seemingly involved, would be to transfer this squeezing to an optical mode. This process
has been considered in Ref.~\cite{aorzsag}, where three electronic levels of an ion trapped inside a cavity would be used to transfer the motional squeezing
to light squeezing of the cavity mode. A fiber collecting the output light from the cavity could be used to transfer this squeezing to the target.

\end{document}